\documentclass[superscriptaddress,preprintnumbers]{revtex4}

\usepackage{amsmath,amssymb,amsfonts}
\usepackage{graphicx}
\usepackage{array}
\usepackage{xcolor}
\usepackage{mathrsfs}

\allowdisplaybreaks

\begin{document}
\title{Physical Features of Geometrically Deformed Anisotropic Charged Three-dimensional BTZ Black Holes}

\author{Z. Yousaf}
\email{zeeshan.math@pu.edu.pk}
\affiliation{Institute of Mathematics, University of the Punjab, Lahore-54590, Pakistan}

\author{Kazuharu Bamba}
\email{bamba@sss.fukushima-u.ac.jp}
\altaffiliation{Corresponding author}
\affiliation{Faculty of Symbiotic Systems Science, Fukushima University, Fukushima 960-1296, Japan}

\author{Mansoor Alshehri}
\email{mhalsheri@ksu.edu.sa}
\affiliation{Department of Mathematics, College of Science, King Saud University, P.O.Box 2455 Riyadh 11451, Saudi Arabia}

\author{S. Khan}
\email{suraj.pu.edu.pk@gmail.com}
\affiliation{Department of Mathematics, Quaid-i-Azam University, 45320, Islamabad 44000, Pakistan}

\author{M. Z. Bhatti}
\email{mzaeem.math@pu.edu.pk}
\affiliation{Institute of Mathematics, University of the Punjab, Lahore-54590, Pakistan}
\affiliation{Research Center of Astrophysics and Cosmology, Khazar University, 41 Mehseti Street, 1096, Baku, Azerbaijan}

\begin{abstract}
This work employs the minimal geometric deformation decoupling scheme to derive interior stellar solutions in the background of an electrically charged BTZ ansatz as a seed metric in three dimensions. In this respect, we impose two different equations of state to determine the deformation function and the new material contributions emerging from the additional field source.
%the components of the $\Theta$-field source.
Furthermore, we describe the finiteness of all thermodynamic quantities of the presented stellar solutions, including the effective thermodynamical quantities, for varying values of the deformation parameter and total electric charge. We explore the new interior astrophysical solutions in three-dimensional gravity by analyzing the charged BTZ metric, admitting circular symmetry through the principles of geometric deformation. This study examines the impact of radial-metric deformation on the charged BTZ geometry and underscores the importance of stellar decoupling within the context of electrically charged dense distributions. It is shown that new physically acceptable solutions by incorporating any known three-dimensional spacetime as the isotropic basis are possible, which in turn enable one to analyze the quantum effects due to low degrees of freedom at lower dimensions.
\end{abstract}
\maketitle

\section{Introduction}

One of the most significant issues in mathematical physics is the coexistence of general relativity (GR) and quantum gravitational physics. A few decades ago, the idea that black holes (BHs) could be essential probes of quantum gravity came to prominence. Hawking radiation and the no-hair theorem are two key properties of these specific GR solutions, which have significantly influenced BH quantum mechanics research. BHs also exist within the less complex domain of three-dimensional/$(2+1)$D gravity, exhibiting many properties similar to their higher-dimensional counterparts. Despite the absence of gravitational waves, $2+1$ dimensional gravity is now known to encode a variety of intriguing features \cite{carlip2003quantum}. For example, Brown and Henneaux \cite{brown1986central} have pioneered the discovery of a centrally extended asymptotic symmetry algebra, along with its Chern-Simons (CS) formulation \cite{achucarro1986chern} in $(2+1)$D gravity.

Understanding the gravitational field through the principles of $(2+1)$D gravity remains an active area of study. The simplicity that results from having fewer propagating degrees of freedom is one argument in favor of concentrating quantum gravity research on smaller dimensions. Moreover, a fundamental feature of gravity in $(2+1)$D is its strong relationship to CS theory
\cite{witten19882,achucarro1986chern,achucarro1989extended,barnich2017geometric,rincon2020anisotropic}. Therefore, considering the significance of $(2+1)$D black hole (BH) models, particularly the widely recognized Ba\~{n}ados-Teitelboim-Zanelli (BTZ) BH solution, which serves as the most iconic example \cite{banados1992black,banados1993geometry}, it is worthwhile to examine and investigate potential modifications of these solutions and related concepts. Another significant point to emphasize, which supports our thirst for knowledge about gravitational interactions in lower dimensions, is its vital role in the anti-de Sitter/conformal field theory correspondence
\cite{maldacena1999large,strominger1998black,balasubramanian1999stress,aharony2000large}.
Therefore, the concept of gravity in $(2+1)$D is important in several ways, as was previously mentioned. One of the key benefits of exploring $(2+1)$D solutions is the ability to completely characterize spacetime for basic topologies. This provides valuable insights into BH physics and the underlying quantum gravitational structure. Another important element of $(2+1)$D gravity theories is their growing importance in condensed matter physics. This interaction provides a platform for research into the development of dense materials that mimic the behavior of BHs. This creates the opportunity to experimentally test concepts emerging from quantum gravity and dualities \cite{franz2018mimicking}. In addition to the well-established BTZ-BH solutions, there are other relativistic stellar solutions such as linear \cite{cataldo1996static}, nonlinear electrodynamics
\cite{cataldo1999three,cataldo2000regular,rincon2017scale,rincon2018quasinormal},
scale-dependent BHs \cite{koch2016scale,rincon2017btz,rincon2018scale}, and numerous
other configurations as detailed \cite{darabi2013generalized,he20172}.

One of the most dynamic fields of stellar physics involves the study of BHs from both observational and theoretical perspectives. BHs are fascinating not only because of their unusual properties but also because they could test GR and offer important new insights into
gravity in strong-field regimes. However, there are many challenges in the way of integrating theoretical predictions and observational evidence. A significant advancement in this area is
the recent direct observation of BHs via the detection of gravitational waves, marking the beginning of a new and promising era for gravitational physics \cite{abbott2016observation,abbott2017gw170104}. A surge of fascination in lower-dimensional gravity was sparked by the discovery of the BTZ-BH solution
in (1+2) dimensions \cite{banados1992black}. Three-dimensional gravity is special due to its strong relationship to the Chern-Simons theory \cite{jackiw1990classical} and its lack of propagating degrees of freedom. In addition, it offers a framework through which we can study a mathematically simpler three-dimensional system to obtain insights into four-dimensional realistic BHs. {The Chern-Simons corrections in the inflationary Lagrangian alter the tensor perturbations and produce a chiral spectrum, but they have no direct impact on the background evolution. In this direction, Nojiri \emph{et al.} \cite{nojiri2020propagation} used Chern–Simons axion gravity to study the development of cosmological gravitational waves. They demonstrated the non-equivalent propagation of these modes, which results in chiral tensor modes. Odintsov and Oikonomou \cite{odintsov2022chirality} investigated the chirality of the produced primordial gravitational waves by taking into account an axionic Chern-Simons-corrected $f(R)$ model. More specifically, they examined two primary axion models that offer an intriguing particle phenomenology with $R^2$ corrections in the inflationary Lagrangian: the kinetic axion model and the classic misaligned axion model. Furthermore, to produce a visible energy spectrum of stochastic primordial gravitational waves consistent with the 2023 NANOGrav measurements, Oikonomou \cite{oikonomou2023flat} presented several examples from particle physics and theories of gravity.} A negative cosmological constant is usually the genesis of the BTZ black hole. However, alternative sources such as scalar fields and electromagnetic fields have also been explored, including scale-dependent variations of these models. The static BTZ-BH is the prototypical BH solution in $(2+1)$D. It is important to note that this solution does not exhibit any inherent pressure anisotropy. Therefore, it would be worthwhile to first explore the incorporation of anisotropies into that geometry. In the context of charged BTZ solution, Peld\`{a}n \cite{peldan1993unification} developed rotationally symmetric and static solutions for the Einstein-Maxwell theory considering a unified gravity model and the Yang-Mills theory.

Therefore, it could potentially be worthwhile to look into how local anisotropy might be incorporated into that geometry initially. One of the successful and simplest theoretical techniques for introducing anisotropies into the relativistic stars and BHs is the
geometric decoupling of stellar distributions via minimal geometric deformation (MGD). This effective technique was initially developed against the backdrop of the Randall-Sundrum brane world. In more challenging circumstances, anisotropic solutions can also be identified within the scope of BH physics. In recent years, GR has witnessed the application of MGD decoupling to extend isotropic fluid solutions to non-isotropic domains
\cite{ovalle2017decoupling,ovalle2018anisotropic,ovalle2019decoupling,khan2025quasi,khan2025zhao}.
Consequently, the MGD method has gained increasing popularity for
obtaining new and physically significant solutions within GR and
other gravitational models. It is important to highlight that local
anisotropy can be introduced into widely recognized spherically
symmetric isotropic stellar models for self-gravitational stars,
yielding feasible approximations of stellar formations. Building on the effectiveness of the approach in $(3+1)$D, it is sensible to study its application in the lowest
dimensional framework consistent with GR: $(2+1)$D. Staruszkiewicz's seminal work asserts that $(2+1)$D gravity is a theory without a gravitational field, and space is flat with no
material content. However, examining the GR-field equations in $(2+1)$D
with matter has yielded valuable insights. These solutions have
served as a testing ground for exploring features similar to those
in $(3+1)$D. They encompass point particle solutions, perfect
fluids, cosmological models, dilatons, inflatons, and
string-inspired scenarios \cite{banados1992black}. For a recent comprehensive review of
$(2+1)$D exact solutions, see \cite{rincon2020anisotropic,contreras2019extended}.

Ever since the introduction of Einstein's relativistic model of
gravitation, GR, it has been quite challenging to determine
solutions that characterize physically acceptable stellar
configurations in the universe from a scientific standpoint. The
understanding of gravity and its influence on the vast and
mysterious stellar structures in our universe are based on the
analytical solutions to GR-field equations. In this direction, the
gravitational field exterior to the spherical fluid sphere was
described by Schwarzschild \cite{schwarzschild1916gravityfield}, who
was the first to obtain an exact solution to the GR-field equations.
Despite the construction of numerous approximate as well as exact
isotropic stellar solutions in GR, most lack physical significance
and fail to meet basic astrophysical observational constraints
\cite{delgaty1998physical}. Theoretical
frameworks offer limited isotropic stellar models under static and
spherically symmetric relativistic constraints. Furthermore, only a
a small fraction of these models exhibit physical feasibility
according to basic astrophysical observations. On the other hand,
the occurrence of certain physical phenomena within stellar
distributions show that no self-gravitating system is constituted
solely of isotropic fluid. This suggests that the assumption of
isotropy for characterizing the fluid content of stellar
distributions may not fully capture the dynamics of compact
entities.

Exploring the dynamics of gravitationally bound systems with
anisotropic matter content has consistently garnered attention and
remains an active area of research due to its physical relevance.
The number of physical phenomena producing anisotropy in both
relativistic and non-relativistic domains have been proved by strong
pieces of evidence \cite{herrera1997local}. A more detailed
understanding of non-isotropic matter configuration was made
possible by Bowers and Liang's study on the equation of state for
locally anisotropic fluid spheres \cite{bowers1974anisotropic}.
Ruderman's studies \cite{ruderman1972pulsars} on more realistic compact, self-gravitational distributions suggest that nuclear matter may transition into an anisotropic regime under an extremely dense state ($\rho > 10^{15 }\textmd{g/cm}^{3}$), necessitating a
relativistic treatment of nuclear interactions. This issue has been
the subject of several recent studies within the domain of GR and
its various modifications, as detailed in
\cite{khan2024structure,khan2024construction,yousaf2024imprints,albalahi2024anisotropic}
and references therein. These studies investigate the influence of
anisotropic fluid content on the physical properties of highly dense
fluid spheres, such as effective mass, radius, central energy
density, critical surface redshift, and stability. In certain cases,
anisotropy introduces a repulsive force (where radial pressure
$P_{r}$ dominates the tangential pressure $P_{\bot}$) that
counterbalances the gravitational pull \cite{mak2003anisotropic}.
However, it is also fascinating to study these anisotropic stellar
fluid spheres in the presence of an electric charge
\cite{albalahi2024electromagnetic,yousaf2024role,albalahi2024isotropization}.
The energy density of the electric field, which greatly affects the
object's gravitational mass is one of the important aspects of
electric charge \cite{deb2018anisotropic}. To understand the effects
of electric charge on relativistic dense stars, numerous studies
\cite{ray2003electrically,negreiros2009electrically,varela2010charged}
can be consulted. Building upon Bonnor's \cite{bonnor1960mass}
pioneering work in general relativity, charged self-gravitational
non-isotropic fluid distributions have been extensively explored.
The interaction of anisotropy and a static electric field could
enhance the equilibrium and stability of highly dense fluid
distributions
\cite{maurya2013charged,maurya2015anisotropic,yousaf2022stability}.

This work is motivated by the increasing interest in how astrophysical stellar configurations behave in low-dimensional spacetime and how geometrical shifts affect their physical characteristics. Investigating the effects of electrical charge on stellar distributions in three-dimensional gravity is critical because it enables us to examine the unique impacts of charge and geometrical deformations on gravitationally confined systems that higher-dimensional models may not entirely describe. We intend to use the minimal geometric deformation scheme of astrophysical decoupling to discover new paths for modeling gravitationally compact, electrical structures, which could shed insight into phenomena such as charged BHs and strange stellar systems. This study makes major contributions to both fundamental physics and the theory of gravitation by establishing an approach for investigating the impact of radial deformations on the geometry of spacetime in lower dimensions. This is especially relevant since quantum effects may become more evident in lower dimensions due to fewer degrees of freedom. Our findings could also be relevant for future studies on the thermodynamics of astrophysical structures and quantum gravity theories. Besides the study of BTZ-BHs, some significant investigations related to the smeared compact objects with different backgrounds must be taken into account \cite{ovgun2016existence,ovgun2021evolving,battista2022gravitational,battista2022propagation,battista2024quantum,capozziello2024avoiding,yang2024black}.

Motivated by the above discussions, we primarily focus in this work
on the geometric deformation of the charged BTZ-BH. This work
derives an anisotropic extension of the charged, BTZ-BH solution by
employing the MGD decoupling scheme. This study differs from previous investigations of minimally and completely deformed versions of BTZ geometry in that it considers the charged BTZ model as the seed metric for constructing both isotropic and anisotropic stellar models. However, in previous research, the vacuum BTZ geometry was extended into the anisotropic domain through gravitational decoupling \cite{contreras2019extended}. The remainder of the article is
organized as follows. Section \textbf{II} summarizes the main features associated with standard GR formalism coupled with additional field sector $\Theta_{\mu\nu}$ in
the $(2+1)$D geometry. In Sec. \textbf{III}, we employ the MGD decoupling scheme to a
static and circularly symmetric self-gravitational source filled
with isotropic fluid matter content. Section \textbf{IV} is
dedicated to generating minimally deformed stellar solutions from
the static and charged BTZ background. Finally, we provide a summary
of our findings in Sec. \textbf{V}.

\section{The General Framework of Gravitational Decoupling}

The extended form of the gravitational action in the context of gravitational decoupling can be expressed as
\begin{align}\label{a1}
S=\int\left[\frac{R}{2\kappa^{2}}+{L}_{m}
+\beta{L}_{\Theta}\right]\sqrt{-g}d^{4}x,
\end{align}
where ${L}_{m}$ and ${L}_{\Theta}$ represent the Lagrangian densities of the matter field and the additional field source beyond GR, respectively, and $R$ is the scalar invariant associated with the Ricci tensor $R_{\mu\nu}$. Furthermore, $\kappa^{2}$ denotes the gravitational coupling constant, and $g=\textmd{det}(g_{\mu\nu})$. The stress-energy tensors corresponding to the matter fields and the additional field source can be written as
\begin{align}\label{a2}
T_{\mu\nu}&=-\frac{2}{\sqrt{-\textsl{g}}}\frac{\delta(\sqrt{-g}{L}_{m})}{\delta g^{\mu\nu}}=-2\frac{\delta({L}_{m})}{\delta g_{\mu\nu}}+g^{\mu\nu}{L}_{m},
\end{align}
and
\begin{align}\label{a4}
\Theta_{\mu\nu}=2\frac{\delta({L}_{\Theta})}{\delta g^{\mu\nu}}-g_{\mu\nu}{L}_{\Theta},
\end{align}
respectively. The resulting expressions for the GR stellar structure equations associated with the action \eqref{a1} are as follows
\begin{align}\label{a5}
G_{\mu\nu}\equiv R_{\mu\nu}-\frac{1}{2}\mathrm{R}g_{\mu\nu}=\kappa^{2}\left
(T^{(m)}_{\mu\nu}+\beta\Theta_{\mu\nu}\right),
\end{align}
which can be rewritten as
\begin{align}\label{a6}
G_{\mu\nu}=\kappa^{2}T_{\mu\nu}^{(tot)}, \quad \textmd{with} \quad T_{\mu\nu}^{(tot)}=T^{(m)}_{\mu\nu}+\beta\Theta_{\mu\nu},
\end{align}
where
\begin{align}\label{a7}
{T^{\mu}_{~\nu}}^{(m)}=\textmd{diag}(-\rho^{(m)},P^{(m)},P^{(m)}),
\end{align}
denotes the stress-energy tensor corresponding to the isotropic fluid, and $\Theta_{\mu\nu}$ is the generic field source that is coupled to ${T^{\mu}_{~\nu}}^{(m)}$ by the parameter $\beta$. The divergence-free feature of the Einstein tensor $G_{\mu\nu}$ implies that the total stress-energy tensor, $T_{\mu\nu}^{(tot)}$, is conserved, as described by the equation
\begin{align}\label{w1}
\nabla^{\mu}T_{\mu\nu}^{(tot)}=0.
\end{align}
The metric ansatz representing a static circularly symmetric stellar configuration reads
\begin{align}\label{a8}
ds^{2}=\textsl{g}_{\mu\nu}dx^{\mu}dx^{\nu}=-e^{\xi}dt^{2}+e^{\eta}dr^{2}+r^{2}d\theta^{2},
\end{align}
where $\xi=\xi(r)$ and $\eta=\eta(r)$, and $x^{\mu}\equiv x^{0,1,2}=(t,r,\theta)$. Then, applying the decoupled stellar equations \eqref{a6} to the metric \eqref{a8}, we obtain
\begin{align}\label{a17}
G^{0}_{~0}=\kappa^{2}{T^{0}_{~0}}^{(tot)}:\kappa^{2}\left({T^{0}_{~0}}^{(m)}+\beta\Theta^{0}_{~0}\right)&=\frac{\eta'e^{-\eta}}{2r},
\\\label{a18}
G^{1}_{~1}=\kappa^{2}{T^{1}_{~1}}^{(tot)}:\kappa^{2}\left({T^{1}_{~1}}^{(m)}+\beta\Theta^{1}_{~1}\right)&=\frac{\xi'e^{-\eta}}{2r},
\\\label{a19}
G^{2}_{~2}=\kappa^{2}{T^{2}_{~2}}^{(tot)}:\kappa^{2}\left({T^{2}_{~2}}^{(m)}+\beta\Theta^{2}_{~2}\right)&=-\frac{e^{-\eta}}{4}
\left(\eta'\xi'-\xi'^{2}-2\xi''\right),
\end{align}
where $f'\equiv\partial_{r}f.$ The straightforward observation of the above-stated system allows us to write
\begin{align}\label{a20}
\rho=&\rho^{(m)}+\beta \Theta^{0}_{~0},
\\\label{a21}
P_{r}=&P^{(m)}+\beta \Theta^{1}_{~1},
\\\label{a22}
P_{t}=&P^{(m)}+\beta \Theta^{2}_{~2}.
\end{align}
On the other hand, the divergence-free expression \eqref{w1} explicitly reads
\begin{align}\label{a23}
P'+\frac{\xi'}{2}(\rho+P)-\beta(\Theta^{1}_{~1})'+\frac{\beta\xi'}{2}\left(\Theta^{0}_{~0}-\Theta^{1}_{~1}\right)
+\frac{\beta}{r}(\Theta^{0}_{~0}-\Theta^{1}_{~1})=0.
\end{align}
Thus, the distribution of the entire stellar structure can be characterized using the stress-energy tensor
\begin{align}\label{a23a}
{T^{\mu}_{~\nu}}^{(\textmd{tot})}=\textmd{diag}\left(-\rho,+P_{r},+P_{t}\right),
\end{align}
where $\rho$, $P_{r}$, and $P_{\bot}$ represent the energy density, radial pressure, and tangential pressure, respectively. It is noted that the stellar structure Eqs. \eqref{a17}--\eqref{a19} describe the Einstein system associated with the anisotropic fluid. Thus, the additional field source $\Theta_{\mu\nu}$ induces pressure anisotropy within the stellar system. The degree of this anisotropy is determined by the deformation parameter $\beta$, which vanishes within the limit $\beta\rightarrow0$. Now, to solve the gravitational system characterized by the set of Eqs. \eqref{a17}--\eqref{a19} and \eqref{a23}, we need to determine the unknowns $\{\eta,\xi,\rho,P_{r},P_{\bot}\}$. One common scheme used for the reduction of unknowns is to introduce an equation of state associating the components of the stress-energy tensor. However, the solutions provided in this paper are found using the MGD scheme, as detailed below.

\section{Geometric Deformation through MGD Scheme}

This section focuses on introducing the decoupling scheme for separating stellar distributions using the radial metric deformation, also known as the MGD approach. This approach allows us to study how the $\Theta$-gravitational fluid source affects the seed metric under consideration. This straightforward geometric deformation can be encoded in the following linear transformation of the radial metric
function
\begin{align}\label{a28}
e^{-\eta(r)}=\lambda(r)+\beta f(r).
\end{align}
The linear map known as MGD decoupling has interesting implications. This map indicates that MGD affects only one metric potential, leaving the other (temporal) component unchanged. Here, the geometric deformation is encoded with the help of the decoupling function $f$. Then, by employing the transformation \eqref{a28} into the stellar structure Eqs. \eqref{a17}--\eqref{a19}, the system is separated into two sets. By setting $\beta$ equal to zero, we obtain one set of solutions that represents an isotropic distribution
\begin{align}\label{a30}
&2\kappa^{2}r\rho=-\lambda',
\\\label{a31}
&2\kappa^{2}rP=\lambda\xi',
\\\label{a32}
&4\kappa^{2}P=\lambda'\xi'+2\lambda\xi''+\lambda\xi'^{2},
\end{align}
whose conservation reads
\begin{align}\label{a33}
P'+\frac{\xi'}{2}(\rho+P)=0.
\end{align}
The second set is associated with the extra field source $\Theta_{\mu\nu}$ and is defined as
\begin{align}\label{a34}
2\kappa^{2}r\Theta^{0}_{~0}=&-f',
\\\label{a35}
2\kappa^{2}r\Theta^{1}_{~1}=&-f\xi',
\\\label{a36}
-4\kappa^{2}\Theta^{2}_{~2}=&f'\xi'+2f\xi''+f\xi'^{2},
\end{align}
with the following conservation equation
\begin{align}\label{a37}
\left(\Theta^{1}_{~1}\right)'-\frac{\xi'}{2}\left(\Theta^{0}_{~0}-\Theta^{1}_{~1}\right)-\frac{1}{r}
\left(\Theta^{2}_{~2}-\Theta^{1}_{~1}\right)=0.
\end{align}
It is significant to notice that the stellar structure equations regulating the anisotropic $\Theta_{\mu\nu}$ and isotropic fluid configurations follow Einstein's field equations, in contrast to the $3+1$ dimensional research discussed in Refs. \cite{ovalle2017decoupling,ovalle2018anisotropic,ovalle2018black,las2018using}. This shows that $G_{\mu\nu}$ is a linear superposition of two individual Einstein tensors, $G_{\mu\nu}^{(m)}$ and $\mathcal{G}_{\mu\nu}^{(m)}$, each satisfying Einstein's field equations independently. Therefore, we have
\begin{align}\label{n1}
G_{\mu\nu}^{(m)}=\kappa^{2}T_{\mu\nu}^{(m)},
\end{align}
for the perfect fluid field source, and
\begin{align}\label{n2}
\mathcal{G}_{\mu\nu}^{(m)}=\kappa^{2}\Theta_{\mu\nu}^{(m)},
\end{align}
for the additional anisotropic field sector. Therefore, we can express
\begin{align}\label{n3}
G_{\mu\nu}=G_{\mu\nu}^{(m)}+\mathcal{G}_{\mu\nu}^{(m)}.
\end{align}
This result can be generalized to any number of gravitational field sources. The Einstein field equations for multiple sources can be decomposed into individual Einstein equations for each source, that is
\begin{align}\label{n4}
T^{(tot)}_{\mu\nu}=T_{\mu\nu}^{(m)}+\sum_{i}\beta_{i}\Theta^{(i)}_{\mu\nu}, \quad i>1,
\end{align}
and
\begin{align}\label{n5}
\nabla^{\mu}T^{(m)}_{\mu\nu}=\nabla^{\mu}\Theta_{\mu\nu}^{(1)}=\cdot\cdot\cdot=\nabla^{\mu}\Theta_{\mu\nu}^{(n)}=0.
\end{align}
The Einstein tensor corresponding to the total stress-energy tensor, $T^{(tot)}_{\mu\nu}$, can be decomposed as follows
\begin{align}\label{n6}
G_{\mu\nu}=G_{\mu\nu}^{(m)}+G_{\mu\nu}^{(1)}+\cdot\cdot\cdot+G_{\mu\nu}^{(n)},
\end{align}
where the tensorial terms are related in the following manner
\begin{align}\label{n7}
G_{\mu\nu}^{(m)} &=\kappa^{2}T^{(m)}_{\mu\nu},
\\\nonumber
G_{\mu\nu}^{(1)} &=\kappa^{2}\beta \Theta^{(1)}_{\mu\nu},
\\\nonumber
\vdots&\quad\quad\vdots
\\\nonumber
G_{\mu\nu}^{(n)} &=\kappa^{2}\beta \Theta^{(n)}_{\mu\nu}.
\end{align}
This conclusion is significant because anisotropic systems in $3+1$ dimensional metric obey quasi-Einstein field equations instead of Einstein field equations, unlike isotropic configurations \cite{ovalle2017decoupling,ovalle2018anisotropic,ovalle2018black}. This discrepancy arises from an absence of the $-1/r^{2}$-term, which prevents the resemblance to classical Einstein equations. Additionally, in $3+1$ dimensions, although the equations exhibit a quasi-Einstein nature, conservation laws can be formulated as a linear combination of these quasi-Einstein equations. Consequently, the isotropic matter configuration and the extra field source $\Theta_{\mu\nu}$ interact exclusively through gravity without exchanging energy. This fact can be outlined by the expression
\begin{align}\label{n8}
\nabla^{\mu}T^{(tot)}_{\mu\nu}=\nabla^{\mu}\left(T^{(m)}_{\mu\nu}+\beta\Theta_{\mu\nu}\right)=0,
\\\nonumber
\Rightarrow\nabla^{\mu}T^{(m)}_{\mu\nu}=\nabla^{\mu}\Theta_{\mu\nu}=0.
\end{align}
This is why, in some studies, the generic field sector $\Theta_{\mu\nu}$ is interpreted as the content of dark matter. The next section applies the MGD scheme to construct new stellar models based on the static, circularly symmetric, electrically charged $2+1$ BTZ spacetime.

\section{Junction Conditions}
In the study of gravitationally confined systems, junction conditions are of vital importance as they provide crucial insights into the interior configuration and underlying physics. In (2+1)D, the boundary is a circumference ($r = R$) that separates the interior and exterior metrics. The interior metric is defined as
\begin{align}\label{aa10}
ds^{2}=-e^{\xi^{-}}dt^{2}+e^{\eta^{-}}dr^{2}+r^{2}d\theta^{2},
\end{align}
and the exterior metric is given by
\begin{align}\label{aa11}
ds^{2}=-e^{\xi^{+}}dt^{2}+e^{\eta^{+}}dr^{2}+r^{2}d\theta^{2}.
\end{align}
The outer functions are solutions to the field equations $G_{\mu\nu} =\Theta_{\mu\nu}$. Furthermore, the anisotropies $\Theta_{\mu\nu}$ are expressed using the classical metric functions, $\xi(r)$ and $\eta(r)$, which are those of the BTZ-BH solution. To smoothly join the two solutions, we impose the continuity of both the first and second fundamental forms. The first fundamental form, when evaluated on the surface, takes the following form
\begin{align}\label{aa12}
\left[ds^{2}\right]_{\Sigma}=0.
\end{align}
This yields two junction conditions
\begin{align}\label{aa13}
e^{\xi^{-}}-e^{\xi^{+}}\Big|_{\Sigma}=0,
\\\label{aa14}
e^{\eta^{-}}-e^{\eta^{+}}\Big|_{\Sigma}=0.
\end{align}
From Eq. \eqref{aa14}, we can write
\begin{align}\label{aa15}
-M+R^{2}-Q^{2}\ln R+\beta f(R)=e^{\xi^{+}(R)},
\end{align}
where $M$ is the mass of and $f$ is the deformation function. The second fundamental form is defined as \cite{contreras2019general}
\begin{align}\label{aa16}
\left[G_{\mu\nu}r^{\mu}\right]_{\Sigma}=0,
\end{align}
where $r^{\mu}$ denotes the radial vector. The condition \eqref{aa16} provides \cite{contreras2019general}
\begin{align}\label{aa17}
P_{r}^{-}-P_{t}^{+}\Big|_{\Sigma}=0.
\end{align}
Lastly, applying the definition of $\Theta^{1}_{~1}$ yields
\begin{align}\label{aa18}
P_{R}=\beta\left[-\frac{v'_{R}f_{R}}{16\pi R}+\frac{g_{R}}{8\pi}\left(-m+R^{2}-\frac{s^{2}}{2}\ln R\right)^{-1}\right],
\end{align}
by using the following relations
\begin{align}\label{aa19}
P_{R}=P(R)^{-},\quad v'_{R}=v'(R)^{-}.
\end{align}
It is also important to note that the deformation function associated with the outer BTZ solution, $g(r)$, is provided by the anisotropic contribution $\Theta_{\mu\nu}$, where the metric retains the same form as in Eq. \eqref{a40}, subject to the appropriate substitution
\begin{align}\label{aa20}
e^{\eta(r)}=-m+r^{2}-Q^{2}\ln r+\beta \textmd{g}.
\end{align}
Thus, the necessary and sufficient conditions for matching the internal MGD metric with the exterior metric, defined by the charged deformed BTZ solution, are given by Eqs. \eqref{aa13}, \eqref{aa14}, and \eqref{aa18}, respectively. However, if we set $\textmd{g}=0$, the metric reduces to the outer charged BTZ solution, given by
\begin{align}\label{aa21}
P_{R}\equiv P(R)+\beta\left(-\frac{v'_{R}f_{R}}{16\pi R}\right)=0,
\end{align}
where $\Theta^{1}_{~1}(R)=-\frac{v'_{R}f_{R}}{16\pi R}$.

\section{Minimally Deformed Charged BTZ Black Hole Solutions}

The coupling of GR and Maxwell field theory in a three-dimensional universe yields the charged BTZ-BH as a solution, which is described in the following action \cite{gwak2016thermodynamics,martinez2000charged}
\begin{align}\label{a38}
S=\int d^{3}x\sqrt{-g}\left(\frac{1}{2\kappa^{2}}(R-2\Lambda)-\frac{1}{4}F_{\mu\nu}F^{\mu\nu}\right)+\beta S_{\Theta},
\end{align}
where the gravitational constant $\kappa^{2}=8\pi G$ is set to be unity and the cosmological constant is $\Lambda$ is considered to be $-1$. Furthermore,
\begin{align}\label{a10}
F_{\mu\nu}:=\nabla_{\mu}A_{\nu}-\nabla_{\nu}A_{\mu},
\end{align}
represents the antisymmetric electromagnetic field tensor and $S_{\Theta}$ denotes the action for an extra gravitational field source. Now, varying the action \eqref{a38} provides the following field equations
\begin{align}\label{as10}
G_{\mu\nu}+\Lambda \textsl{g}_{\mu\nu}&=\kappa^{2}(T_{\mu\nu}+\beta\Theta_{\mu\nu}),
\\\label{zx10}
\nabla_{\mu}F^{\mu\nu}&=0,
\end{align}
where $\Theta_{\mu\nu}$ comes from the variation of $S_{\Theta}$.
The stress-energy tensor corresponding to the electromagnetic field is defined as
\begin{align}\label{a13}
 T_{\mu\nu}=F_{\mu\alpha}F_{\nu}^{~\alpha}-\frac{1}{4}\textsl{g}_{\mu\nu}F_{\alpha\gamma}F^{\alpha\gamma}.
\end{align}
 The standard metric in the
context of $(2+1)$D-BH solutions can be expressed as follows
\begin{align}\label{a39}
ds^{2}=g_{\mu\nu}dx^{\mu}dx^{\mu}=-e^{\xi(r)}dt^{2}+e^{\eta(r)}dr^{2}+r^{2}d\theta^{2}.
\end{align}
The simplest form, satisfying to the standard Reissner-Nordstr\"{o}m condition, is the well-known charged BTZ solution, characterized by a lapse function of
\begin{align}\label{a40}
e^{\xi(r)}=\lambda(r)=-m+r^{2}-Q^{2}\ln r,
\end{align}
with an electric charge $Q$, given as
\begin{align}\label{az40}
A_{\mu}=-Q\ln r dt.
\end{align}
The values of the physical variables associated with the above system are given by
\begin{align}\label{w42}
\rho=&\frac{Q^{2}-2r^{2}}{2r^{2}},
\\\label{w43}
P=&-\frac{Q^{2}-2r^{2}}{2r^{2}}.
\end{align}
The static BTZ solution is known to represent a black hole in a spacetime with a cosmological constant. Now, we will use the MGD-decoupling approach to modify the BTZ black hole solution. More specifically, we will introduce a specific field source,
$\Theta_{\mu\nu}$, to fill the spacetime. This will result in a deformed BTZ geometry.
The minimally deformed version of the metric \eqref{a39} can be given by
\begin{align}\label{a41}
ds^{2}=-e^{\xi(r)}dt^{2}+\frac{dr^{2}}{\lambda(r)+\beta f(r)}+r^{2}d\theta^{2}.
\end{align}
The values of the non-zero $\Theta$-components corresponding to $A_{\mu}$ \eqref{az40} and the deformed BTZ metric \eqref{a41} are defined as
\begin{align}\label{z42}
\Theta^{0}_{~0}=&\frac{1}{2r^{2}h}\left[rhf'+Q^{2}f\right],
\\\label{z43}
\Theta^{1}_{~1}=&\frac{f}{h},
\\\label{z44}
\Theta^{2}_{~2}=&\frac{1}{4r^{2}h^{2}}\left[-(Q-2r^{2})rhf'+f\{-4Q^{2}r^{2}\ln r-4r^{2}(m-Q^{2})-Q^{4}\}\right].
\end{align}

%\begin{figure}[H]
%\centering{{\includegraphics[height=2.3 in, width=3.4 in]{Pic.eps}}}
%\caption{Circularly symmetric geometry containing additional field
%source $\Theta_{\mu\nu}$. Notably,the case $\Theta_{\mu\nu}=0$
%provides the $(2+1)$D charged BTZ-BH.}\label{a1}
%\end{figure}

\subsection{Isotropic Solutions}

To close the stellar system described by the set of Eqs. \eqref{a17}--\eqref{a19}, we need additional information. Theoretically, we can either choose any expression for the deformation parameter $f(r)$ that yields a physically acceptable solution or apply restrictions to $\Theta_{\mu\nu}$ to achieve the desired outcome. By imposing an isotropic pressure constraint on the additional field source $\Theta_{\mu\nu}$, as in \cite{ovalle2018black}, we find
\begin{align}\label{a41a}
\Theta^{1}_{~1}=\Theta^{2}_{~2},
\end{align}
which on substituting the values of $\Theta^{1}_{~1}$ and $\Theta^{2}_{~2}$ from Eqs. \eqref{a34} and \eqref{a35}, provides
\begin{align}\label{a43}
f=\left(Q^{2}-2r^{2}\right)\left[rhf'+\left(Q^{2}-2r^{2}\right)\right]=0,
\end{align}
which is a first-order ODE, whose integration yields the following deformation function
\begin{align}\label{v43}
f=f_{0}\left(-m+r^{2}-Q^{2}\ln r\right),
\end{align}
where $f_{0}$ is an integration constant.
Now, the modified form of the radial metric function turns out to be
\begin{align}\label{a44}
e^{-\eta(r)}=\left(1+\beta f_{0}\right)(-m+r^{2}-Q^{2}\ln r).
\end{align}
The above result explains how the standard charged BTZ metric is modified by gravitational decoupling via MGD. The addition of factor $(1+\beta f_{0})$ introduces a scaling correction to the original solution, thereby changing the causal structure of the spacetime, which may be interpreted as the result of modifications from an alternative gravity theory or the influence of an additional matter source. Finally, the minimally deformed form of the metric \eqref{a41} turns out to be
\begin{align}\label{d44}
ds^{2}=\left(-m+r^{2}-Q^{2}\ln r\right)dt^{2}-\frac{dr^{2}}{\left(1+\beta f_{0}\right)(-m+r^{2}-Q^{2}\ln r)}-r^{2}d\theta^{2}.
\end{align}
Then, the values of the thermodynamical quantities can be expressed as
\begin{align}\label{a45}
\rho&=\frac{(Q^{2}-2r^{2})(1+f_{0}\beta)}{2r^{2}},
\\\label{a46}
P_{r}&=P_{t}=P=-\frac{(Q^{2}-2r^{2})(1+f_{0}\beta)}{2r^{2}}.
\end{align}
Importantly, the set of Eqs. \eqref{a45} and \eqref{a46}
characterize an isotropic stellar distribution within the formalism
of $(2+1)$D model. The values of the Ricci scalar $\mathrm{R}$ and
Kretschmann scalar $\mathrm{K}$ the  after simple calculations turn
out to be
\begin{align}\label{z1}
R&=\frac{(Q^{2}-6r^{2})(1+\beta f_{0})}{r^{2}},
\\\label{z2}
\mathrm{K}&=\frac{\left(3Q^{2}-4Q^{2}r^{2}+12r^{4}\right)(1+\beta f_{0})^{2}}{r^{4}}.
\end{align}
The expressions of both the curvature scalars show that there is a curvature singularity at $r=0$. It is important to note that the charged BTZ-BH solution can be recovered within the limit $\beta\rightarrow0$. The horizons occur where the metric potential
$e^{-\eta(r)}=0$, which gives
\begin{align}\label{aa1}
\left(1+\beta f_{0}\right)(-m+r^{2}-Q^{2}\ln r)=0,
\end{align}
which gives $m=r_{h}^{2}-Q^{2}\ln r_{h}\equiv m_{h}(r_{h})$ (for $1+\beta f_{0}\neq=0$), $r_{h}$ being the radius of the Killing horizon, where
\begin{align}\label{az2}
r_{ex}\equiv \frac{Q}{\sqrt{2}}; \quad m_{ex}:=\frac{Q^{2}}{2}\left[1-\ln\left(\frac{Q^{2}}{2}\right)\right],
\end{align}
which shows that $r=r_{ex}$ is a degenerate horizon. In the case where $m> m_{ex}$, the spacetime exhibits two non-degenerate Killing horizons. When $m= m_{ex}$, a single degenerate Killing horizon is present. For $m < m_{ex}$ no Killing horizons exist. On the other hand, the particular case where
$1+\beta f_{0}=0$ must be taken into account. In this case, the metric function $e^{-\eta(r)}$ is identically zero, implying that every radial value represents a horizon. This would be an extremely unusual and likely nonphysical outcome.

\subsection{Conformally Symmetric Stellar Solutions}

For this case, we consider the consider the following constraint
\begin{align}\label{a47}
\Theta^{0}_{~0}+\Theta^{1}_{~1}+\Theta^{2}_{~2}=0,
\end{align}
indicating a traceless source, consistent with conformal symmetry \cite{sato2022conformally}.
Then using the values of the $\Theta$-components, $\Theta^{0}_{~0}$, $\Theta^{1}_{~1}$ and $\Theta^{2}_{~2}$ from Eqs. \eqref{a34}--\eqref{a36}, we get
\begin{align}\label{a48}
-&\left[Q^{2}(Q^{2}+4r^{2})\ln r+2r^{4}+r^{2}(4m-3Q^{2})+Q^{2}\left(m+\frac{Q^{2}}{2}\right)\right]f
\\\label{vv48}
-&r(-m+r^{2}-Q^{2}\ln r)\left(Q^{2}\ln r+\frac{Q^{2}}{2}-2r^{2}+m\right)f'=0,
\end{align}
whose solution cannot be obtained in a closed form. Further development of these types of solution can be done in future studies.

At this stage, some significant remarks are in order:
\begin{itemize}
    \item It is significant to note that the authors of Refs. \cite{rincon2020anisotropic} developed the exterior
BTZ-BH solution within the formalism of gravitational decoupling.
However, our work is related to extending this scheme to the
interior solution describing the evolution of isotropic astrophysical models corresponding to the charged BTZ
metric, which has not been explored previously. Specifically, the
investigations performed in Refs.
\cite{rincon2020anisotropic} describes the
charged BTZ metric in a barotropic or polytropic framework, focusing
on exterior BH solutions. However, by focusing on the
internal structure of the charged BTZ solution, our study fills a
major gap. It has important consequences for comprehending the
behavior of matter and its distribution in a lower-dimensional
charged environment.
    \item We utilized the MGD decoupling method to develop the charged interior BTZ
    solution, expanding its application beyond exterior charged
    configurations. Our extension offers novel perspectives on the nature of interior stellar models in the
    $(2+1)$-D formalism. Specifically, we discuss how the charge and the gravitational field interact inside the star or compact object.
    In lower-dimensional spacetimes, this inner solution can offer a thorough understanding of the thermal characteristics of charged entities and gravitational collapse.
    \item Previous studies using gravitational decoupling to derive charged solutions have primarily focused on the exterior
     vacuum behavior of the charged BTZ solution. We analyze the solutions acquired in our study to those presented in Refs.
\cite{rincon2020anisotropic},
     showing how our results reinforce the external solutions.
    \item Our approach has considerable physical consequences, especially when dealing with compact objects in a lower-dimensional charged context.
    The charged internal stellar solution provides a new perspective on how charge affects the inner framework, stability, and probable equilibrium configurations.
    This unexplored element may have implications for comprehending strange star objects in lower-dimensional gravity models.
  \end{itemize}

\subsection{Behavior of thermodynamical quantities}

\begin{figure}[htbp]
\centering
\begin{minipage}{0.48\linewidth}
  \centering
  \includegraphics[width=\linewidth]{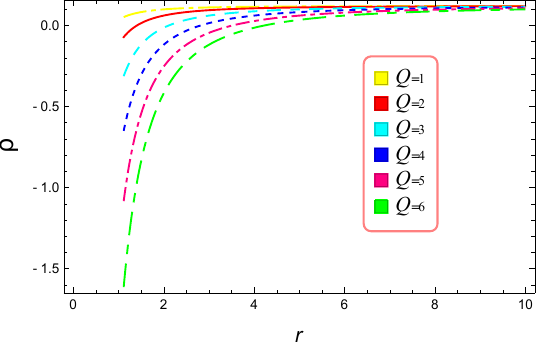}
\end{minipage}\hfill
\begin{minipage}{0.48\linewidth}
  \centering
  \includegraphics[width=\linewidth]{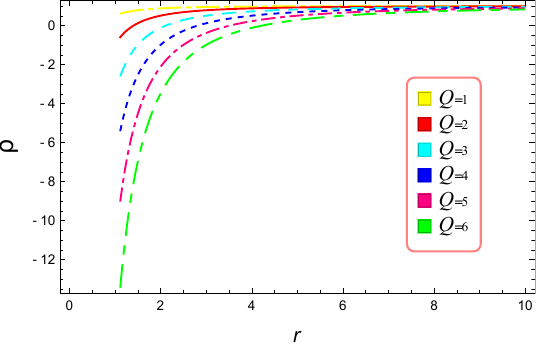}
\end{minipage}
\caption{The behavior of $\rho$ is shown for $\beta = 1$, with $f_{0} = -1.2$ (left panel) and $f_{0} = -2$ (right panel), for $Q=1$ (yellow dash-dotted line), $Q=2$ (red solid line), $Q=3$ (cyan dashed line), $Q=4$ (blue dotted line), $Q=5$ (pink dash-dotted line), and $Q=6$ (green space-dashed line).}
\label{1f}
\end{figure}

\begin{figure}[htbp]
\centering
\begin{minipage}{0.48\linewidth}
  \centering
  \includegraphics[width=\linewidth]{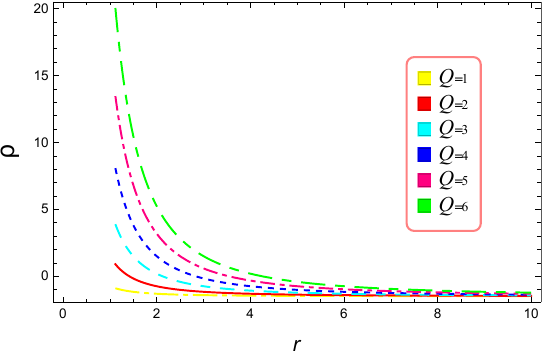}
\end{minipage}\hfill
\begin{minipage}{0.48\linewidth}
  \centering
  \includegraphics[width=\linewidth]{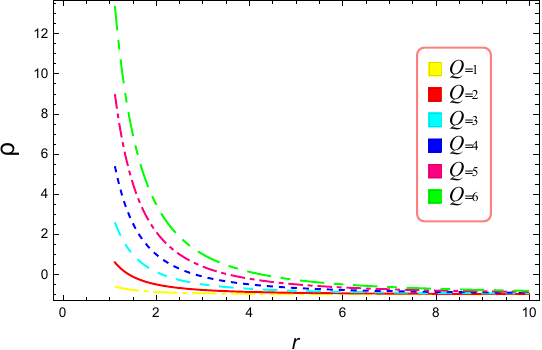}
\end{minipage}
\caption{The variation of $\rho$ for $\beta=1$, $f_{0}=0.5$ (left panel) and for $f_{0}=0$ (right panel) for different values of $Q$. The legend is the same as in Fig.~\ref{1f}.}
\label{2f}
\end{figure}

\begin{figure}[htbp]
\centering
\begin{minipage}{0.48\linewidth}
  \centering
  \includegraphics[width=\linewidth]{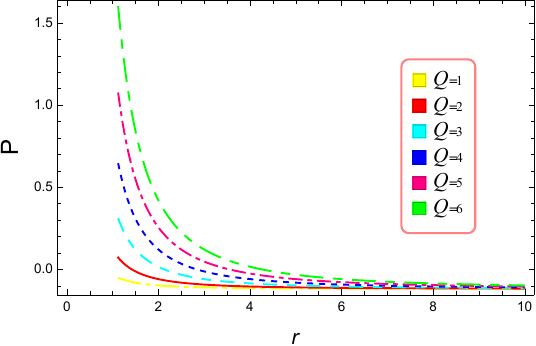}
\end{minipage}\hfill
\begin{minipage}{0.48\linewidth}
  \centering
  \includegraphics[width=\linewidth]{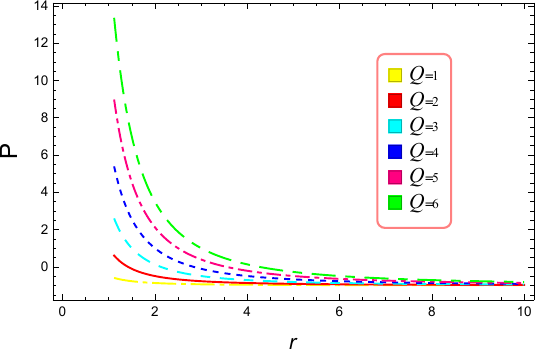}
\end{minipage}
\caption{The behavior of $P$ for $\beta=1$, $f_{0}=-1.2$ (left panel) and for $f_{0}=-2$ (right panel) for different values of $Q$. The legend is the same as in Fig.~\ref{1f}.}
\label{3f}
\end{figure}

\begin{figure}[htbp]
\centering
\begin{minipage}{0.48\linewidth}
  \centering
  \includegraphics[width=\linewidth]{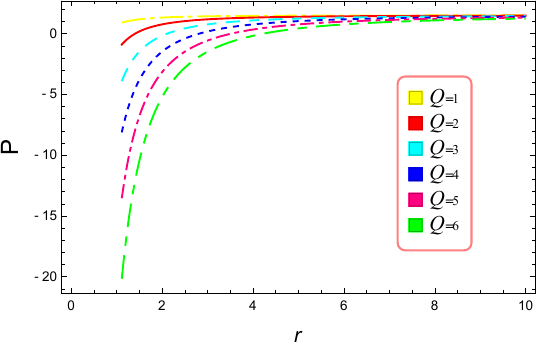}
\end{minipage}\hfill
\begin{minipage}{0.48\linewidth}
  \centering
  \includegraphics[width=\linewidth]{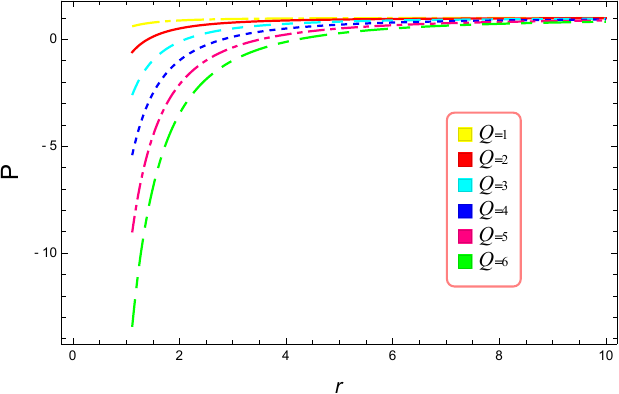}
\end{minipage}
\caption{The variation of $P$ for $\beta=1$, $f_{0}=0.5$ (left panel) and for $f_{0}=0$ (right panel) for different values of $Q$. The legend is the same as in Fig.~\ref{1f}.}
\label{4f}
\end{figure}

The behavior of the thermodynamical quantities $\rho$ and $P$ associated with the charged BTZ model under conformal symmetry, considering $\beta=1$ and the parameter values $f_{0}=-1.2$, $f_{0}=-2$, $f_{0}=0.5$, and $f_{0}=0$, is described
in Figs. \ref{1f}--\ref{4f}.The thermodynamic quantities must be positive, finite, and monotonically decreasing. For $Q=1$,
 $Q=2$, $Q=2$, $Q=3$, $Q=4$, $Q=5$ and $Q=6$, the density and pressure exhibit monotonically decreasing behavior.
 
%\begin{figure}[H]
%\centering{{\includegraphics[height=2.5 in, width=3.0 in]{Rho.eps}}}
%\centering{{\includegraphics[height=2.5 in, width=3.0 in]{Rho1.eps}}}
%\caption{The behavior of $\rho$ is shown for $\beta = 1$, with $f_{0} = -1.2$ (left panel) and $f_{0} = -2$ (right panel), for $Q=1$ (yellow dash-dotted line), $Q=2$ (red solid line), $Q=3$ (cyan dashed line), $Q=4$ (blue dotted line), $Q=5$ (pink dash-dotted line), and $Q=6$ (green space-dashed line).}\label{1f}
%\end{figure}
%\begin{figure}[H]
%\centering{{\includegraphics[height=2.5 in, width=3.0 in]{Rho2.eps}}}
%\centering{{\includegraphics[height=2.5 in, width=3.0 in]{Rho3.eps}}}
%\caption{The variation of $\rho$ for $\beta=1$, $f_{0}=0.5$ (left panel) and for $f_{0}=0$ (right panel) for different values of $Q$. The legend is the same as in Fig. \ref{1f}.}\label{2f}
%\end{figure}
%\begin{figure}[H]
%\centering{{\includegraphics[height=2.5 in, width=3.0 in]{P1.eps}}}
%\centering{{\includegraphics[height=2.5 in, width=3.0 in]{P2.eps}}}
%\caption{The behavior of $P$ for $\beta=1$, $f_{0}=-1.2$ (left panel) and for $f_{0}=-2$ (right panel) for different values of $Q$. The legend is the same as in Fig. \ref{1f}.}\label{3f}
%\end{figure}
%\begin{figure}[H]
%\centering{{\includegraphics[height=2.5 in, width=3.0 in]{P3.eps}}}
%\centering{{\includegraphics[height=2.5 in, width=3.0 in]{P4.eps}}}
%\caption{The variation of $P$ for $\beta=1$, $f_{0}=0.5$ (left panel) and for $f_{0}=0$ (right panel) for different values of $Q$. The legend is the same as in Fig. \ref{1f}.}\label{4f}
%\end{figure}

\subsubsection{Energy conditions}

\begin{figure}[htbp]
\centering
\includegraphics[width=0.6\linewidth]{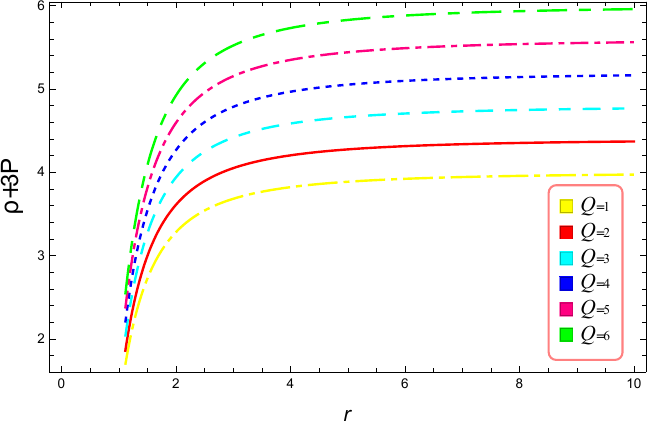}
\caption{Behavior of SEC for $\beta=1$ and $f_{0}=1$ for different values of $Q$.
The legend is the same as in Fig.~\ref{1f}.}
\label{5f}
\end{figure}

This section will explore a specific set of constraints imposed on the considered solution. To ensure the physical plausibility of the solution, we establish restrictions on the physical parameters known as energy conditions. These energy constraints are categorized as null (NEC), weak (WEC), strong (SEC), and dominant (DEC). These constraints are defined by the following set of mathematical relations
\begin{align}\label{w1a}
&\textmd{NEC}:\quad\rho+P\geq0,
\\\label{w2}
&\textmd{WEC}:\quad\rho\geq0,~\rho+P\geq0,
\\\label{w3}
&\textmd{SEC}:\quad\rho+3P\geq0,~\rho+P\geq0,
\\\label{w4}
&\textmd{DEC}:\quad\rho\geq0,~\rho\geq|P|.
\end{align}
The energy conditions can be interpreted in the following way:
Figs. \ref{1f}--\ref{4f} illustrate that NEC, WEC, and DEC are satisfied. The behavior of SEC is shown in Fig. \ref{5f}. As a result, we have a well-behaved energy-momentum tensor.

From a physical perspective, the NEC, WEC, and DEC ensure that the
energy density perceived by any observer remains positive. Moreover,
the satisfaction of the SEC guarantees the attractive nature of
gravity. Figure \ref{5f} shows that the
SEC is satisfied at all points, indicating that the core of the celestial body has a well-defined and positive stress-energy tensor. The graphical analysis shows that the decoupling parameter $\beta$ directly influences the astrophysical characteristics described by the above inequalities, making them more positive for higher $\beta$ values. Table~\ref{tab:tableS3} describes the summary of the physical variables and the energy conditions corresponding to the proposed model.

\subsection{Speed of Sound}

Since for the relativistic solutions, we have $\rho=-P$. Therefore, the speed of sound is defined as
\begin{align}\label{ss1}
c_{s}^2=\frac{d\rho}{dP}=-1,
\end{align}
That is the speed of sound is negative. This shows that the system exhibits a Jeans instability, resulting in the exponential growth of perturbations rather than oscillatory behavior.
\begin{table}[htbp]
\caption{\label{tab:tableS3}%
Summary of the astrophysical features associated with the
charged BTZ solution.}
{\begin{ruledtabular}
\begin{tabular}{cccccccc}
Thermodynamic variable & Charged BTZ solution
%& Graphical behavior
& Figures
& Features\\
\hline
 Density ($\rho$) & $\beta=1$,~$f_{0}=-1.2$ &
%$\rho$ versus $r$
Fig. 1 (left panel)
& smoothly decreasing\\
 $\rho$ & $\beta=1$,~$f_{0}=-2$ &
%$P_{r}$ versus $r$
Fig. 1 (right panel)
& smoothly decreasing \\
 $\rho$ & $\beta=1$,~$f_{0}=0.5$ &
%$P_{t}$ versus $r$
Fig. 2 (left panel)
& smoothly decreasing \\
 $\rho$ & $\beta=1$,~$f_{0}=0$ &
%$\Pi$ versus $r$
Fig. 2 (right panel)
& smoothly decreasing  \\
Pressure (P) & $\beta=1$,~$f_{0}=-1.2$ &
%$\rho$ versus $r$
Fig. 3 (left panel)
& smoothly decreasing\\
 P & $\beta=1$,~$f_{0}=-2$ &
%$P_{r}$ versus $r$
Fig. 3 (right panel)
& smoothly decreasing \\
 P & $\beta=1$,~$f_{0}=0.5$ &
%$P_{t}$ versus $r$
Fig. 4 (left panel)
& smoothly decreasing \\
 P & $\beta=1$,~$f_{0}=0$ &
%$\Pi$ versus $r$
Fig. 4 (right panel)
& smoothly decreasing  \\
 NEC (Null Energy Condition) &
 $\rho+P\geq0$ &
%NEC versus $r$
Figs. 1 to 4
& satisfied  \\
 WEC (Weak Energy Condition) & $\rho\geq0,~\rho+P\geq0$ &
%WEC versus $r$
Figs. 1 to 4
& satisfied \\
 DEC (Dominant Energy Condition) & $\rho\geq0,~\rho\geq|P|$ &
%DEC versus $r$
Figs. 1 to 4
& satisfied \\
SEC (Strong Energy Condition) &$\rho+3P\geq0,~\rho+P\geq0$ &
%SEC versus $r$
Fig. 5
& satisfied \\
\end{tabular}
\end{ruledtabular}}
\end{table}

%\section{Summary and Discussions}
\section{Conclusions}

In this investigation, we applied the MGD form of decoupling stellar sources to $(2+1)$D static and circularly symmetric geometry. We obtained both the isotropic and anisotropic minimally deformed solutions that model a self-gravitational star embedded in charged BTZ-BH. Unfortunately, the conformally flat solutions are not obtained in closed form due to the complexity of the system.  Our model shows that the isotropic sector obeys the standard GR structure, unlike the $3+1$ dimensional case, where the anisotropic field source corresponds to the quasi-Einstein system. Within this framework, the GR field equations governing a system of stellar sources can be split into independent field equations corresponding to each distinct field
sector. To demonstrate the method, we obtained new stellar solutions by applying the geometric deformation approach to the well-known static and electrically charged BTZ geometry. More precisely, the isotropic and conformal equations of state were utilized to solve the anisotropic gravitational source $\Theta_{\mu\nu}$. The presented results mimic the $(3+1)$D charged version of the investigation \cite{ovalle2018black}. In particular, the MGD version
of geometric deformation gives rise to new structures with higher curvatures including causal horizons and singularities. The presented results offer a versatile framework for constructing new solutions by incorporating any known $(2+1)$D spacetime as the isotropic basis and at the same time enable us to analyze the quantum effects due to low degrees of freedom at lower dimensions. As a specific application, we showed that the approach may be used to generate regular BH solutions and derive novel $(2+1)$D solutions that generalize the well-known charged BTZ metric.

It is important to point out that even though isotropic perfect fluid compact configurations may not represent the most precise approximations of the real astrophysical distributions, the approach still offers fruitful insights due to several advantages. By expanding the collection of known isotropic solutions, isotropic solutions can contribute to the construction of more realistic stellar models. Additionally, perfect fluid configurations can precisely describe phenomena like the initial states of a realistic star. Moreover, these solutions can be applied to model cold, fluid planetary bodies \cite{rahman2002spacetime,martin2004algorithmic,hansraj2019impact,aoki2020minimally}. The isotropic matter content is also helpful in characterizing the configuration of charged self-gravitational dense stars \cite{maharaj2007generalized}. This new solution thus contributes to the restricted number of previously known theoretically feasible isotropic systems.
The main goal of this research is to determine how gravity, electromagnetism, and quantum effects interact in lower-dimensional spacetimes. By investigating the inner solutions of electrically charged gravitational configurations in three dimensions, we aim to obtain insight into the fundamental nature of gravitational interactions and possible quantum corrections. Furthermore, this study presents an innovative technique for studying the impact of geometric deformations on the physical properties of compact structures, which may provide insight into matter's behavior under lower-dimensional complex mechanisms. These studies are critical to improving our understanding of BH physics, quantum gravity, and the early cosmos.

This study could be extended to incorporate additional field contributions, such as exotic matter sources and scalar fields like dilaton fields, to examine their impact on stellar configurations. The consequences of these solutions could be analyzed, particularly for charged BTZ-like structures in lower-dimensional gravity, within the framework of the AdS/CFT correspondence. Furthermore, this investigation could be applied to exploring quantum gravitational effects using the gravitational decoupling scheme.

\renewcommand{\theequation}{\arabic{equation}}
\setcounter{equation}{0}

%\section*{Appendix}

\vspace{0.3cm}

%%%%%%%%%%%%%%%%%%%%%%%%
%%%  Acknowledgments
%%%%%%%%%%%%%%%%%%%%%%%%

\vspace{0.3cm}

\section*{Acknowledgement}

The work of KB was supported in part by the JSPS KAKENHI Grant Numbers 24KF0100, 25KF0176 and Competitive Research Funds for Fukushima University Faculty (25RK011). The work of MA was supported by the Researchers Supporting Project number: ORF-2026-411, King Saud University, Riyadh, Saudi Arabia.

\section*{Conflict of Interest}

The authors declare no conflict of interest.

\section*{Data Availability Statement}

This manuscript has no associated data or the data will not be deposited. [Authors comment: This manuscript contains no associated data.]

\end{document}